\documentstyle[aps,twocolumn]{revtex}

\newcommand{\bq}{\begin{equation}}
\newcommand{\eq}{\end{equation}}
\newcommand{\bqn}{\begin{eqnarray}}
\newcommand{\eqn}{\end{eqnarray}}
\newcommand{\nb}{\nonumber}
\newcommand{\lb}{\label} 

\begin{document} 

\draft

\title{Instability of cosmological event horizons of nonstatic 
global cosmic strings II: perturbations of gravitational waves and
massless scalar field }
\author{ Jos\'e A.C. Nogales \thanks{e-mail address: 
 jnogales@portela.if.uff.br}}
\address{Instituto de F\' {\i}sica, Universidade Federal Fluminense,
Av. Litor\^anea s$/$n - Boa Viagen,  CEP 24210-340  Niter\'oi~--~RJ, 
Brazil\\
and Instituto de F\' {\i}sica, Universidad Mayor de San Andres (UMSA), 
Casilla 3553,  La Paz~--~Bolivia }
\author{Anzhong Wang \thanks{e-mail address: wang@symbcomp.uerj.br}}
\address{ Departamento de F\' {\i}sica Te\' orica,
Universidade do Estado do Rio de Janeiro, 
Rua S\~ ao Francisco Xavier 524, Maracan\~ a,
20550-013 Rio de Janeiro~--~RJ, Brazil }

\date{ November 21, 1997 }
\maketitle

\begin{abstract}
The stability of the cosmological event horizons (CEHs) of a class of
non-static global cosmic strings is studied against perturbations of
gravitational waves and massless scalar field. It is found that the
perturbations of gravitational waves always turn the CEHs into
non-scalar weak spacetime curvature singularities, while the ones of
massless scalar field turn the CEHs either into non-scalar weak
singularities or into scalar ones depending on the particular cases
considered. The perturbations of test massless scalar field is also
studied, and it is found that they do not always give the correct
prediction.
\end{abstract}


\section{INTRODUCTION}

Cosmic strings which may have been formed in the early Universe have
been studied extensively \cite{VS1994}, since the pioneering work of
Kibble \cite{Kibble1976}. Recently, Banerjee {\em et al.}
\cite{Banerjee1996} and Gregory \cite{Gregory1996} studied non-static
global strings, and some interested results were found. In particular,
Gregory showed that the spacetime singularities usually appearing in the
static case \cite{HS1988} can be replaced by cosmological event horizons
(CEH). This result is very important, as it may make the structure
formation scenario of cosmic strings more likely, and may open a new
avenue to the study of global strings.

However, our recent studies \cite{WN1997} showed that these
CEHs in general were not stable to the perturbations of null dust fluid,
and always turned into spacetime singularities. The singularities
are strong in the sense that the distortion of the test particles
diverges logarithmically.

In this paper, we shall study the stability of the CEHs against
perturbations of massless scalar field  and gravitational waves.
Specifically, the paper is organized as follows: in the Sec. II we
consider the perturbations of a test massless scalar field, while in
Sec.III and Sec. VI,  we  consider the ``physical" perturbations of
gravitational waves and massless scalar field, respectively. The word
``physical" here means that the back reaction of the perturbations is
taken into account. The paper is closed by Sec. V, where our main
conclusions are derived. 

The main purpose of studying perturbations of test massless scalar field
is to generalize the Helliwell-Konkowski (HK) conjecture about the
stability of quasi-regular singularities \cite{HK1992} to the stability
of CEHs. As a matter of fact, in \cite{WN1997} which will be referred as
Paper I, it was shown that the conjecture works well and gives the
correct predictions about the stability of the CEHs, as far as the
perturbations of null dust fluid are concerned.

The notations used in this paper will closely follow the ones used in
Paper I,  and to avoid of repeating, some results given there will be
directly used without any further explanations.

\section{THE PERTURBATIONS OF TEST MASSLESS SCALAR FIELDS}
 
\renewcommand{\theequation}{2.\arabic{equation}}
\setcounter{equation}{0}

By requiring that the string has fixed proper width and that the
spacetime has boost symmetry in the ($t, z$)-plane, Gregory managed to
show that the spacetime for a U(1) global string (vortex) is given by
the metric \cite{Gregory1996}
\bqn
\lb{2.1}
ds^{2} &=& e^{2 A(r)}dt^{2} - dr^{2} - e^{2[A(r) + b(t)]}dz^{2}\nb\\ 
& & - C^{2}(r)d\theta^{2}.
\eqn
For the cases where $b(t) =  \ln[\cosh(\beta t)], \; + \beta t,\; -
\beta t$, with $\beta$ being a positive constant, the metric
coefficients inside the core of a string have the asymptotic behavior
\bq
\lb{2.2}
e^{A(r)} \sim \beta(r_{0} - r),\;\;\; C(r) \sim C_{0} + O(r_{0} -
r)^{2}, 
\eq 
as $r \rightarrow r_{0}^{-}$, where $C_{0}$ is a constant [cf. Eq.(3.14)
in Ref. 5]. It was shown that in all the three cases the hypersurface $r
= r_{0}$ represent a cone-like CEH \cite{Gregory1996,WN1997}.  

To study the stability of these CEHs against perturbations of massless
scalar field and gravitational waves, it is found convenient to introduce
 two null coordinates, $u$ and $v$, via the relations,
\bq
\lb{2.3}
u=\frac{t + R}{\sqrt{2}}, \;\;\;\;  v =\frac{t - R}{\sqrt{2}},
\eq
where 
\bq
\lb{2.4}
R \equiv \int{e^{-A(r)} dr} =  
- \frac{1}{\beta}\ln[\beta(r_{0} - r)]. 
\eq

In terms of $u$ and $v$, the Gregory solutions
can be cast in the form,
\bqn
\lb{2.5}
ds^{2}&=& 2e^{-M_{(0)}}du dv \nb\\
& & - e^{-U_{(0)}} \left[e^{V_{(0)}}
dz^{2} + e^{-V_{(0)}}d\theta^{2}\right],
\eqn
Where $M_{(0)} = \sqrt{2}\beta (u - v)$, and 
\bq
\lb{2.6}
U_{(0)} = \left\{ \begin{array}{ll}
 -  \ln\left[\cosh\left(\frac{\beta}
{\sqrt{2}}(u + v)\right)\right] & \\
+ \frac{\beta}{\sqrt{2}}(u - v) - \ln c_{0}, & b(t) = 
 \ln[cosh(\beta t)],\\
- \sqrt{2}\beta v - \ln C_{0}, & b(t) = \beta t,\\
\sqrt{2}\beta u - \ln C_{0}, & b(t) = - \beta t.
\end{array} \right.
\eq
Note that in Paper I, the perturbations were considered in both regions
$r \le r_{0}$ and $r \ge r_{0}$. However, as shown there, the
conclusions obtained in these two regions  are the same. Thus,  without
loss of generality, in the rest of the paper, we shall restrict
ourselves only to the region $r \le r_{0}$. Then, the Klein-Gordon
equation $\phi,_{\mu}\phi,_{\nu} g^{\mu\nu} = 0$ takes the form
\bq
\lb{2.7}
2\phi_{,u v} - U_{(0),v}\phi_{,u} + U_{(0),u}\phi_{,v} = 0,
\eq
where $()_{,x} \equiv \partial ()/\partial x$. 

To study the above equation, let us first consider the  case $b(t) = +
\beta t$.  In this case, it can be shown that Eq.(\ref{2.7}) has the
general solution
\bq
\lb{2.8}
\phi(u,v) = F(u) e^{(\alpha u -\beta v)/\sqrt{2}} + G(v),
\;\; [b(t) = + \beta t],
\eq
where $F(u)$ and $G(u)$ are arbitrary functions of their indicated
arguments.  To have the perturbation be finite initially ($t = -
\infty$), we require that the two arbitrary functions be finite as $t
\rightarrow - \infty$ and $\alpha \ge \beta$. Then, the trace of the
energy-momentum tensor (EMT) $T_{\mu\nu}$ for the test massless scalar
field is given by
\bqn
\lb{2.9}
T &=& T^{\lambda}_{\lambda} = - \phi_{,\mu}\phi_{,\nu}g^{\mu\nu}\nb\\
&=& - 2\left[F'(u) + \frac{\alpha F(u)}{\sqrt{2}}\right] 
e^{[\alpha u + \beta(2u -3v)]/\sqrt{2}} \nb\\
& & \times \left[G'(v) - \frac{\beta F(u)}{\sqrt{2}}
e^{(\alpha u - \beta v)/\sqrt{2}} \right],
\eqn
which diverges as $r \rightarrow r^{-}_{0}$, where a prime denotes the
ordinary derivative with respect to their indicated arguments.
Therefore, when the back reaction of the perturbations is taken into
account, we would expect that the CEH will be turned into scalar curvature
singularity, provided that the HK conjecture continuously holds for CEHs
\cite{HK1992}. 

Similarly, it can be shown that the same conclusion is also true for the
case $b(t) = - \beta t$.

When $b(t)=\ln [\cosh(\beta t)]$, Eq.(\ref{2.7}) has the general
solutions
\bq
\lb{2.10}
\phi(u,v) = \sum_{n} \frac{ b_{n}e^{\beta (u-v)/\sqrt{2}}}
{ \left[(a_{n}e^{ \sqrt{2}\beta u} +2)
(a_{n}e^{-\sqrt{2}\beta v}-2)\right]^{1/2}},
\eq
where $\{b_{n}\}$ and $\{a_{n}\}$ are integration constants. Projecting
the corresponding EMT onto the PPON frame defined by Eqs.(A3) and (A4)
in Paper I, we find that the non-vanishing components are given by
\bqn
\lb{2.11}
T_{(0)(0)} &=& T_{(1)(1)}= C^{2}_{+}\phi^{2}_{,u}  +
 C^{2}_{-}\phi^{2}_{,v}\nb\\
T_{(0)(1)} &=& C^{2}_{+}\phi^{2}_{,u}  - 
C^{2}_{-}\phi^{2}_{,v}\nb\\
T_{(2)(2)} &=& T_{(3)(3)}= e^{2\beta R} \phi_{,u} \phi_{,v}, 
\eqn
where
\bqn
\lb{2.12}
C_{\pm} &\equiv&  \frac{1}{\sqrt{2} \beta^{2}(r_{0} - r)^{2}}\left\{
E \pm \epsilon [E^{2} - \beta^{2}(r_{0} - r)^{2}]^{1/2}\right\},\nb\\
\phi_{,u} &=& \sqrt{2} \beta \sum^{\infty}_{n = 1}{\frac{b_{n}e^{\beta R}
[e^{\beta(R - t)} - 2]}{[a_{n}^{2}e^{2\beta R} - 4\sinh\beta t e^{\beta R}
- 4]^{3/2}}},\nb\\
\phi_{,v} &=& \sqrt{2} \beta \sum^{\infty}_{n = 1}{\frac{b_{n}e^{\beta R}
[e^{\beta(R + t)} - 2]}{[a_{n}^{2}e^{2\beta R} - 4\sinh\beta t e^{\beta R}
- 4]^{3/2}}},
\eqn
where $E$ is a constant.  From the above expressions we can see that, as
$t \rightarrow - \infty$, these tetrad components vanish, and as $R
\rightarrow + \infty \; (r \rightarrow r_{0})$, the components
$T_{(0)(0)},\; T_{(1)(1)}$, and $ T_{(0)(1)} $ become unbounded, while
the ones $T_{(2)(2)}$  and $ T_{(3)(3)}$ remain finite.  Thus, after the
back reaction of the perturbations of the massless scalar field is taken
into account, we would expect that the CEH is turned into a spacetime
curvature singularity. However, unlike the last two cases, the nature of
the singularity should be a non-scalar one, since now all the scalars
built from $T_{\mu\nu}$ are finite, for example,
\bqn
\lb{2.13}
T = T^{\lambda}_{\lambda} &=& - 
\frac{1}{2}e^{2\beta R} \phi_{,u} \phi_{,v} \sim Const.\nb\\
  T^{\lambda \delta} T_{\lambda\delta} &=& 
\frac{1}{4}e^{4\beta R} \phi^{2}_{,u} \phi^{2}_{,v} \sim Const.
\eqn
as $ R \rightarrow + \infty$.
To verify whether or not the above analysis gives the correct prediction
for the stability of the CEHs, let us turn to consider real
perturbations, that is, taking the back reaction of the perturbations into
account.

\section{ PERTURBATIONS OF GRAVITATIONAL WAVES}

\renewcommand{\theequation}{3.\arabic{equation}}
\setcounter{equation}{0}

In Paper I, it was noted that, although the study of test null dust fluid
and the one of real null dust fluid all gave the same results on the
instability of the CEHs, the cause of the instability was different.
For the real perturbations, it was caused by the
non-linear interaction of gravitational waves, rather than what the
study of the test particles indicated that they should be caused by the back
reaction of perturbations of null dust fluids. Thus, to study the role
that gravitational waves can play, we devote this section to
perturbations of pure gravitational waves. These perturbations are
always expected to exist, since at the time when the strings were formed,
the temperature of the Universe was very high, and the spacetime was
filled with gravitational and particle radiation \cite{VS1994}.

To study the general perturbation of gravitational waves, it is found
difficult.  In the following we shall study some particular cases.
This does not lose any generality, since if the CEHs are stable, they
should be stable against any kind of perturbations. Otherwise, they are
not stable. Then, from  \cite{W1993} we can easily construct the
following solutions to the Einstein vacuum field equations,
\bq
\lb{3.1}
ds^{2}= 2e^{- M}du dv -e^{-U} (e^{V} dz^{2}+e^{-V} d\theta^{2} ),
\eq
where the metric coefficients are given by
\bqn
\lb{3.2}
M &=& -\ln [a'(u)  b'(v)] - \delta [a(u)-b(v)]\nb\\
   & &  - \frac{\delta^{2}}{4} [a(u)+b(v)]^{2}+ M_{c}, \nb\\
V &=& \ln [a(u)+b(v)] + \delta [a(u)-b(v)] - 2 \ln C_{0},\nb\\
U &=& -\ln [a(u)+b(v)],
\eqn 
where $a(u)$ and $b(v)$ are arbitrary functions, and $\delta, C_{0}$ and
$M_{c}$ are constants. The corresponding Kretschmann scalar is given by
\bqn
\lb{3.3}
{\cal{R}} &\equiv& R_{\alpha\beta\gamma\sigma}R^{\alpha\beta\gamma\sigma}
\nb\\
&=& \delta^{4} [12-\delta^{2}[a(u)+b(v)]^{2} ] \nb\\
& & \times e^{-2 [\delta (a - b)+\delta^{2}(a + b)^{2}/4 - M_{c} ]}.
\eqn
Choosing the null tetrad,
\bqn
\lb{3.4}
l_{\mu} &=& e^{M/2}\delta^{u}_{\mu},\;\; 
n_{\mu} = e^{M/2}\delta^{v}_{\mu},\nb\\
m_{\mu} &=& e^{-U/2}\left[e^{V/2}\delta^{z}_{\mu}
+ i e^{-V/2}\delta^{\varphi}_{\mu}\right],\nb\\
\bar{m}_{\mu} &=& e^{-U/2}\left[e^{V/2}\delta^{z}_{\mu}
- i e^{-V/2}\delta^{\varphi}_{\mu}\right],
\eqn
we find that the non-vanishing components of the Weyl tensor 
$C_{\mu\nu\lambda\sigma}$ are given by
\bqn
\lb{3.5}
\Psi_{0} &=& - C_{\mu\nu\lambda\delta}l^{\mu}m^{\nu}
l^{\lambda}m^{\delta}\nb\\
&= & \frac{\delta^{2}b'(v)^{2}e^{M}}{4} 
\left\{3 - \delta [a(u)+b(v)]\right\}], \nb\\
\Psi_{2} &=& - \frac{1}{2}C_{\mu\nu\lambda\delta}
\left[l^{\mu}n^{\nu}l^{\lambda}n^{\delta} -
l^{\mu}n^{\nu}{m}^{\lambda}\bar{m}^{\delta}\right]\nb\\
&=& -\frac{\delta^{2}e^{M}}{4} a'(u)b'(v),\nb\\
\Psi_{4} &=& - C_{\mu\nu\lambda\delta}n^{\mu}\bar{m}^{\nu}
n^{\lambda}\bar{m}^{\delta}\nb\\
&=& \frac{\delta^{2}a'(u)^{2}e^{M}}{4}
\left\{3 + \delta [a(u)+b(v)]\right\}.
\eqn
The reason to project the Weyl tensor to the null tetrad is that now all
the components $\Psi_{A}$ have their direct physical interpretations
\cite{S1965,W1991}: $\Psi_{0}$  represents the transverse gravitational
wave component along the null direction $l_{\mu}$, $\Psi_{2}$  the
Coulomb-like component, and $\Psi_{4}$ the  transverse gravitational
wave component along the null direction $n_{\mu}$. Since $l_{\mu}$
($n_{\mu}$) defines the outgoing (ingoing) null geodesics \cite{LW1994},
$\Psi_{0}$ ($\Psi_{4}$) now represents the outgoing (ingoing)
cylindrical gravitational wave component.

To use solutions (\ref{3.2}) as the perturbations of gravitational waves
to the Gregory solution, we have to recover them  under certain limits.
To find such limits, let us study the three cases $b(t) =
\ln[\cosh(\beta t)],\; + \beta t, \; - \beta t$ separately.

{\bf a) $b(t) = \beta t$:} In this case, if we choose 
\bq
\lb{3.6}
b(v) = Co e^{\sqrt{2}\beta v}, 
\eq
and replace the null coordinate $u$ by $u'$, where $du' =
e^{-\sqrt{2}\beta u} du$, it can be shown that the solutions given by
Eq.(\ref{3.2}) reduce to the corresponding Gregory solution, as $\delta,
a(u') \rightarrow 0$.   Submitting Eq.(\ref{3.6}) into Eqs. (\ref{3.3})
and (\ref{3.5}), we find that the Kretschmann  scalar and the
$\Psi_{A}$'s are all finites as $t \rightarrow - \infty$, while  near
the CEH where $r \rightarrow r_{0}^{-}$, the Kretschmann scalar
${\cal{R}}$ and the components $\Psi_{0}$ and $\Psi_{2}$ are finite, but
$\Psi_{4}$ becomes infinite. It can be shown that now all the fourteen
scalars built from the Riemann tensor are finite  as $r \rightarrow
r_{0}^{-}$. Therefore, the perturbations of the gravitational waves in
this case do not turn the CEH into a scalar singularity, although they
do turn it into a non-scalar one. The latter can be seen by considering
the tidal forces, represented by the tetrad components of the Riemann
tensor in a free-falling frame (PPON). For example, the component
$R_{(1)(2)(1)(2)}$ in the PPON frame defined by Eqs.(A3) and (A4) in
Paper I diverges like,
\bq
\lb{3.7}
R_{(1)(2)(1)(2)} \rightarrow  (r - r^{-}_{0})^{-2},
\eq
as $r \rightarrow r_{0}^{-}$. Therefore,  the perturbations due to the
gravitational waves turn the CEHs into non-scalar curvature
singularities. However, different from the perturbations of null dust
fluids \cite{WN1997}, now the singularity is weak in the sense that the
distortion, which is equal to the twice integral of the tidal forces,
is finite as $ r \rightarrow r^{-}_{0}$,
\bq
\lb{3.8}
\int\int{R_{(1)(2)(1)(2)}} d\tau d\tau  \rightarrow  (\tau_{0} - \tau)
\ln(\tau_{0} - \tau),  
\eq
where $\tau_{0}$ is a constant and chosen such that $\tau \rightarrow
\tau_{0}$ as $ r\rightarrow r^{-}_{0}$.

It should be noted that in using the PPON frame defined in Paper I to
obtain the above expressions, we have assumed that the gravitational wave
perturbations are weak, so the PPON frame of the perturbed solutions
can be replaced by the one of non-perturbed solutions. This is the case
when $\delta, a(u')$ and its derivatives are all very small. In the
following, whenever we use this frame, we always assume that the
corresponding  conditions hold.

{\bf b) $b(t) = - \beta t$:} In this case, to have the solutions given
by Eq.(\ref{3.2}) reduce to the corresponding Gregory solution, as
$\delta, b(v) \rightarrow 0$, we have to choose 
\bq
\lb{3.9}
a(u) = C_{0}e^{-\sqrt{2}\beta u}.
\eq
Once this is done, it can be shown from Eqs.(\ref{3.3}) and (\ref{3.5})
that now the CEH is also not stable and turned into a non-scalar
singularity  in a manner quite similar to that in the last case. In
particular,  the project of the Riemann tensor onto the PPON frame
defined in Paper I diverges, for example,  the component
$R_{(1)(2)(1)(2)}$ diverges exactly like that of Eq.(\ref{3.7}), while
the twice integral of it is given by Eq.(\ref{3.8}). Thus,  
 the non-scalar singularity is also weak.

{\bf c) $b(t) = \ln[\cosh \beta t]$:} In this case, it is easy to show
that as $\delta \rightarrow 0$, the solutions given by Eq.(\ref{3.2})
reduce to the corresponding Gregory solution, provided the functions
$a(u)$ and $b(v)$ are chosen such that
\bq
\lb{3.12}
a(u)=\frac{C_{0}}{2}e^{-\sqrt{2}\beta u}, \;\;
b(v)=\frac{C_{0}}{2}e^{\sqrt{2}\beta v}. 
\eq
Inserting the above expressions into Eqs.(\ref{3.3}) and (\ref{3.5}) we
find that all of the fourteen scalars built from the Riemann tensor are
finite  both at the initial $t = - \infty$ and as $r \rightarrow
r^{-}_{0}$.  Thus, similar to the last two cases, the perturbations of
the gravitational waves do not turn the CEH into a scalar singularity.
To see whether or not they produce  non-scalar singularities, we can
project the Riemann tensor onto the PPON frame defined in Paper I.
After doing so, we find that some of the tetrad components indeed
diverge, for example, the component $R_{(1)(2)(1)(2)}$ diverges exactly
like that given by Eq.(\ref{3.7}).

Therefore, in all the three cases the gravitational perturbations turn
the CEHs into spacetime singularities, and the singularities are
non-scalar ones, and are weak in the sense that
although the tidal forces diverge, the distortion is finite.

\section{ PERTURBATIONS OF MASSLESS SCALAR FIELD}

\renewcommand{\theequation}{4.\arabic{equation}}
\setcounter{equation}{0}

To study perturbations of massless scalar field, we shall use a theorem
given in \cite{TW1991}, which states as follows: If the solutions
$\left\{M_{g}, U_{g}, V_{g}\right\}$ is a solution of the Einstein {\em
vacuum} field equations for the metric (\ref{3.1}), then, the solution
\bq
\lb{4.1}
\{ M,U,V, \phi \} = \{ M_{g} -\Omega_{g}, V_{g},U_{g},
\lambda V_{g}/\sqrt{2} \},
\eq
is a solution of the Einstein-scalar field equations $G_{\mu\nu} =
\phi_{,\mu} \phi_{,\nu} - g_{\mu\nu}\phi_{,\alpha} \phi^{,\alpha}/2$,
where $\lambda$ is a constant, and
\bq
\lb{4.2}
\Omega_{g} (u,v) = \lambda^{2} \left\{ \frac{3}{2}U_{g}
- \ln |2 U_{g,u} U_{g,v}| - M_{g} \right\} .
\eq
For more details we refer readers to \cite{TW1991}. 

In order to use this theorem, the condition $U_{g,u} U_{g,v} \not= 0$
has to be true.  However, from Eq.(\ref{2.6}) we can see that this is
the case only for $b(t) = \ln[\cosh(\beta t)]$. To overcome this
problem, we shall use the solutions given by Eq.(\ref{3.2}) with $\delta
= 0$ as the vacuum solutions for the cases $b(t) = \pm \beta t$. It can
be shown that in these two cases the corresponding solutions are flat,
and can be brought to the forms that the corresponding Gregory solutions
take by some coordinate transformations. Once this is clear, we take the
solutions given by Eq.(\ref{3.2}) with $\delta = 0$ as the vacuum
solution $\{M_{g}, U_{g}, V_{g}\}$ of the Einstein field equations.
Submitting them into Eq.(\ref{4.1}), we find
\bqn
\lb{4.3}
M &=& (1+\lambda^{2})M_{g} + 
\lambda^{2} \ln 2\left|a'(u)b'(v)\right|\nb\\ 
& & -\frac{\lambda^{2}}{2} \ln[a(u) + b(v)],  \nb\\
V &=& \ln[a(u) + b(v)]  - 2\ln C_{0},\nb\\
U &=& - \ln[a(u) + b(v)],\nb\\
\phi &=& \frac{\lambda}{\sqrt{2}} \left\{\ln[a(u) + b(v)]
   - 2\ln C_{0} \right\},
\eqn
where for the case $b(t) = + \beta t$, the function $b(v)$ is given by
Eq.(\ref{3.6}), while the function $a(u)$ is arbitrary. For the case
$b(t) = - \beta t$, the function $a(u)$ is given by Eq.(\ref{3.9}),
while the function $b(v)$ is arbitrary. For the case $b(t) =
\ln[\cosh(\beta t)]$, the two functions $a(u)$ and $b(v)$ are all fixed
and given by Eq.(\ref{3.12}).  To consider the solutions given by
Eq.(\ref{4.3}) as perturbations of the corresponding Gregory solutions,
we require that the constant $\lambda$, the arbitrary function $a(u)$
and its derivatives in the case $b(t) = + \beta t$, and the arbitrary
function $b(v)$ and its derivatives in the case $b(t) = - \beta t$, are
all small. In particular, when $\lambda,\; a(u) \rightarrow 0$, these
solutions reduce to the Gregory solution for $b(t) = + \beta t$, and
when $\lambda,\; b(v) \rightarrow 0$, they reduce to the Gregory
solution for $b(t) = - \beta t$. 

From Eq.(\ref{4.3}),  
we find that the corresponding physical quantities are given by
\bqn
\lb{4.4}
T &=& T^{\lambda}_{\lambda} = - 
\frac{c_{1}}{[a(u)+b(v)]^{2+\lambda^{2}/2}},\nb\\
{\cal R} &=& \frac{3c_{1}^{2}}{\lambda^{2}
[a(u)+b(v)]^{\lambda^{2}+4}},\nb\\
\Psi_{0} &=& \frac{c_{1}b'(v)}
{4a'(u)[a(u)+b(v)]^{2+\lambda^{2}/2}},\nb\\
\Psi_{2} &=& \frac{c_{1}}
{12[a(u)+b(v)]^{2 + \lambda^{2}/2}}, \nb\\
\Psi_{4} &=& \frac{c_{1}a'(u)}{4b'(v)
[a(u) + b(v)]^{2+\lambda^{2}/2}},
\eqn
where $c_{1}=\lambda^{2}2^{\lambda^{2}}e^{M_{c}(1+\lambda^{2})}$.
To study the asymptotic behavior of these quantities, let us consider
the three cases separately.

{\bf a) $b(t) = \beta t$:} In this case, the function $a(u)$ is
arbitrary but small, and the function $b(v)$ is given by Eq.(\ref{3.6}),
from which we find that
\bq
\lb{4.5}
b(v), \;\; b'(v) \sim  \beta(r_{0} - r)e^{\beta t}.
\eq
Submitting the above expression into Eq.(\ref{4.4}), we find that all
these quantities are finite, except for $\Psi_{4}$ which diverges like
$e^{- \beta t}/(r_{0} - r)$ both as $t \rightarrow - \infty$ and as $r
\rightarrow r^{-}_{0}$.  Note that the amplitude of the gravitational
wave components $\Psi_{0}$ and $\Psi_{4}$ is not completely fixed
\footnote{The amplitude of the two null vectors $l^{\mu}$ and $n^{\mu}$
defined in Eq.(\ref{3.4}) are not completely fixed by the conditions
$l^{\lambda}l_{\lambda} = n^{\lambda}n_{\lambda} = 0$ and
$n^{\lambda}l_{\lambda} = 1$. In \cite{W1991}, it was shown that in
general they take the form $l^{\mu} = B \delta^{\mu}_{v}$ and $n^{\mu} =
A \delta^{\mu}_{u}$, where $AB = e^{M}$. Then, it was found that
$\Psi_{0} = B^{2}\Psi^{(0)}_{0},\; \Psi_{2} = AB\Psi^{(0)}_{2}$, and
$\Psi_{4} = A^{2}\Psi^{(0)}_{4}$, where $\Psi^{(0)}_{A}$'s are
independent of the choice of the functions $A$ and $B$.}. Thus, the
divergence of $\Psi_{4}$ does not really mean that the spacetime is
singular. To clarify this point, let us first consider the fourteen
scalars built from the Riemann tensor, which are found finite for $t = -
\infty$ and $r = r_{0}$.  Therefore, in this case the spacetime is free
of scalar curvature singularities both at the initial and on the CEH. To
see if there exist non-scalar singularities, let consider the tidal
forces. Projecting the corresponding Riemann tensor onto the PPON frame
defined in Paper I, we find that some of its components are diverge, for
example, the component $R_{(1)(2)(1)(2)}$ diverges like Eq.(\ref{3.7}),
while the corresponding distortion  vanishes like that of
Eq.(\ref{3.8}).  Therefore, in this case the perturbations of the
massless scalar  field  turn the CEH into a weak and non-scalar
spacetime singularity.

{\bf b) $b(t) = - \beta t$:} In this case, the function $b(v)$ is
arbitrary and small, and the function $a(u)$ is given by Eq.(\ref{3.8}),
from which we find that
\bq
\lb{4.6}
a(u), \;\; a'(u) \sim \beta(r_{0} - r)e^{- \beta t} \rightarrow \infty,  
\eq
as $t \rightarrow - \infty$, and 
\bq 
\lb{4.7} 
a(u), \;\; a'(u) \sim \beta(r_{0} - r)e^{- \beta t} \rightarrow 0,
\eq
as $r \rightarrow r^{-}_{0}$. Substituting Eq.(\ref{4.6}) into
Eq.(\ref{4.4}) we find that all these quantities are finite as $t
\rightarrow - \infty$. That is, the spacetime is initially free of
spacetime singularities.  However, the combination of
Eqs.(\ref{4.4}) and (\ref{4.7}) shows that the spacetime may be singular
on the CEH $r = r_{0}$. A closer study shows that this is indeed the
case, and similar to the last case, the nature of singularity is a
weak and non-scalar one. 

{\bf d) $b(t) = \ln[\cosh(\beta t)]$:} In this case the two functions
$a(u)$ and $b(v)$ are given by Eq.(\ref{3.12}). Combining this equation
with Eq.(\ref{4.4}) we find that all the quantities given by
Eq.(\ref{4.4}) are finite at the initial, and diverge as $r\rightarrow
r_{0}$.  That is, in this case the CEH is turned into a scalar curvature
singularity.

\section{CONCLUSIONS}

The stability of the CEHs appearing in Gregory's non-static global
cosmic strings \cite{Gregory1996} have been studied. It has been shown
that the gravitational wave perturbations always turn the CEHs into
non-scalar weak spacetime curvature singularities, where ``weak" means that
although the tidal forces become unbounded, the distortion remains
finite as these singularities approach. It has been also shown that the
CEHs are not stable against perturbations of massless scalar fields,
and are turned into non-scalar weak singularities for the cases $b(t) =
\pm \beta t$ and into scalar ones for the case $ b(t) = \ln[\cosh(\beta
t)]$. These results are not consistent with the ones obtained by
studying the perturbations of the test massless scalar field. In
particular, in the last case the latter predicted that the
singularity should be a non-scalar one.  Thus, to generalize the HK
conjecture \cite{HK1992} to the study of the stability of CEHs, more
labor is required.

\section*{ACKNOWLEDGMENTS}

The financial assistance from CNPq (AW), CLAF-CNPq
(JACN), and UMSA (JACN), is gratefully acknowledged.

\end{document}